\documentclass{PoS}

\title{SU(2,CMB), the nature of light and accelerated cosmological expansion }

\ShortTitle{SU(2,CMB), the nature of light and accelerated cosmological expansion }

\author{\speaker{Ralf Hofmann}\\

        Institut f\"ur Theoretische Physik, Universit\"at Frankfurt, 
	Johann Wolfgang Goethe - Universit\"at, 60054 Frankfurt, Germany\\

E-mail: \email{r.hofmann@thphys.uni-heidelberg.de}}

\abstract{We present quantitative and qualitative arguments in favor of the claim that, 
within the present cosmological epoch, the U(1)$_Y$ factor in the 
Standard Model is an effective manifestation of 
SU(2) pure gauge dynamics of Yang-Mills scale $\Lambda\sim 10^{-4}$\,eV. 
Results for the pressure and 
the energy density in the deconfining phase of 
this theory, obtained in a nonperturbative and analytical way, 
support this connection in view of large-angle 
features inherent in the map of the CMB temperature 
fluctuations and temperature-polarization cross correlations.}

\dedicated{Dedicated to Pierre van Baal with best wishes for a 
soon recuperation.}

\FullConference{29th Johns Hopkins Workshop on current problems 
in particle theory: strong matter in the heavens\\

                 1-3 August\\

                 Budapest}

\begin{document}

The principle of relativity and the constancy 
of the velocity of light are the two empirical facts 
Special Relativity is built on. For the electrodynamics of 
moving bodies they imply the nonexistence of a 
singled-out inertial frame of reference: all inertial 
frames are connected by Lorentz transformations under 
which electric and magnetic fields behave like the components 
of a second-rank tensor. As a consequence, 
the long-nurtured idea of a world ether, allegedly enabling 
the propagation of electromagnetic waves and thus defining a 
preferred rest frame, is abandoned \cite{Einstein1905}. 
At first sight this seems to clash wih the observation of an 
almost perfect black-body spectrum in the 
cosmic microwave background (CMB) since one is tempted to tie the 
temperature of the latter to the existence of the 
rest frame of a heat bath. This apparent contradiction 
can, however, be avoided 
if todays CMB and tiny 
deviations in its Planck spectrum by photon emission from 
localized sources (stars), are shown to {\sl decouple} from an existing, 
overall rest frame of a heat bath. 

Conventionally, the 
decoupling of the CMB from its heat bath is 
thought to occur when the Universe becomes 
transparent through the capture of electrons by ions 
and the subsequent formation of neutral atoms. That is, 
at redshift $z\sim 1100$ or temperatures 
in the eV range. This point of view, however, is likely to be overly 
simplistic. As we will argue below, the existence of the CMB is tied 
to the existence of a ground state (or heat bath) which 
only within the {\sl present} cosmological epoch 
happens to be undetectable in the properties of its (photon) excitations. 
(Gravity, however, is sensitive to the 
existence of such an invisible ground state.) 
In other words, the observed decoupling of the background 
radiation from its 
heat bath (ground state) today, implying 
the exact Lorentz covariance of the laws 
of electromagnetism, could be a singled out 
situation in the cosmological evolution. If that is to be the case 
then there must be a good, that is, a dynamical 
reason.      

The purpose of this presentation is to propose a scenario where today's 
Lorentz invariance and certain properties of the CMB, as observed in its 
power spectra at large angles, 
{\sl emerge} due to strongly interacting SU(2) 
Yang-Mills dynamics of scale $\Lambda\sim 10^{-4}$\,eV. We will show 
why the (finite) temperature, where the thermalized SU(2) Yang-Mills theory dynamically restores 
Lorentz invariance, happens 
to be that of the cosmic microwave background $T_{\tiny\mbox{CMB}}=2.35\times 10^{-4}$\,eV. 
Because the scale $\Lambda$ of this Yang-Mills theory essentially is 
$T_{\tiny\mbox{CMB}}$ we adopt the name SU(2)$_{\tiny\mbox{CMB}}$. 

We would like to remark at this point 
that besides the particular situation at $T_{\tiny\mbox{CMB}}$ 
and for $T\ll T_{\tiny\mbox{CMB}}$ \cite{Hofmann2005} 
the Lorentz invariance of the fundamental Lagrangian 
of the SU(2) Yang-Mills theory is an exact symmetry 
of Nature only in the limit of an asymptotically large 
temperature: In this limit all excitations 
are massless, and the ground state, although far from being trivial, neither is 
visible in the spectrum of excitations nor directly 
contributes to any thermodynamical quantity \cite{Hofmann2005}.  

The outline of the presentation is as follows: First, we list a number of 
motivations for the claim that $SU(2)_{{\tiny\mbox{CMB}}}\stackrel{\tiny\mbox{today}}=U(1)_Y$. 
Before we discuss some of the 
physics of the CMB, as it follows from that 
claim, we need to provide prerequisites on a number of 
results, obtained in a nonperturbative and analytical way, 
for the thermodynamics of an SU(2) Yang-Mills theory 
\cite{Hofmann2005,HerbstHofmann2004,HerbstHofmannRohrer2004}. 
Subsequently, we discuss this theory at a particular point $T_{c,E}$ of its 
phase diagram: the boundary between the deconfining 
and preconfining phase. While the ground state of 
the former phase emerges as a spatial average over interacting, topology changing 
quantum fluctuations (calorons and anticalorons subject to 
gluon exchanges between and radiative corrections 
within them which manifest themselves 
in terms of an inert adjoint Higgs field with $T$-dependent modulus on the one hand 
and a pure-gauge configuration on the other hand) the ground state of the
preconfining phase is a condensate of magnetically 
charged monopoles. 

The point $T_{c,E}$ is remarkable because a 
coincidence between an electric and magnetic 
description takes place. (To avoid confusion: A magnetic charge emerging as a result of the apparent 
gauge-symmetry breaking SU(2)$\to$U(1) in the deconfining phase 
is interpreted as an electric charge with respect to U(1)$_Y$. Nevertheless we 
will in the following refer to electric and magnetic 
charges as if the $F_{0i}$ components of the field strength in the 
underlying SU(2) theory defined the color {\sl electric} field.) While 
electrically charged gauge modes decouple 
thermodynamically at $T_{c,E}$ because their mass diverges 
there the charge and the mass of a magnetic monopole 
vanishes at $T_{c,E}$. Thus the photon is precisely 
massless and unscreened at 
$T_{c,E}=T_{\tiny\mbox{CMB}}$: a result which is in agreement 
with our daily experience. This situation is singled-out in the cosmological evolution. 
As a function of temperature a dynamically emerging Lorentz invariance is 
stabilized at $T_{\tiny\mbox{CMB}}$ by an infinitely sharp dip in the energy density. 
Next we investigate to 
what extent the energy density of the ground state at $T_{\tiny\mbox{CMB}}$ and the associated 
(negative) pressure contribute to the Universe's present equation of state. 

A discussion of radiative corrections to the pressure at the 
two-loop level is performed subsequently. We observe 
that within an error of about 50\% the corresponding 
maximal deviation from the pressure of a free photon 
gas at $T\sim 3\,T_{\tiny\mbox{CMB}}$ matches the strength 
of the dipole temperature fluctuation extracted from the CMB map by WMAP and COBE. 
We subsequently discuss this result. 

Finally, we present a summary and 
an outlook on future research. We stress the 
necessity to observationally and theoretically 
investigate the cross correlation 
between electric/magnetic polarization and 
temperature fluctuation at large angles: Information about both 
correlations would allow to identify 
CP violation in the CMB. The identification 
of CP violation, in turn, would support the claim that cosmic coincidence, namely 
the approximate equality of today's energy densities 
in dark matter and dark energy, is explained 
by the slow-roll of a Planck-scale axion 
\cite{Hofmann2005,Wilczek2004}. 
The latter field also would have played 
an important role in the generation of the 
lepton and baryon asymmetries 
as we observe them today \cite{Hofmann2005}.   

\section{Why $SU(2)_{{\tiny\mbox{CMB}}}\stackrel{\tiny\mbox{today}}=U(1)_Y$?}

There are several reasons to consider the possibility that a larger gauge symmetry masquerades as 
the U(1)$_Y$ factor of the standard model within the present cosmological epoch. 
On the theoretical side, Quantum Electrodynamics exhibits ultraviolet slavery as 
opposed to asymptotic freedom - an esthetically not overly appealing property. 
On the observational side there are a number of puzzling large-angle 
anomalies in the one-year data on the cosmic microwave sky as released by 
the WMAP collaboration which, however, have to be 
viewed with a healthy scepticism \cite{Copi2005}. 
Decisive results are expected within the near future. If the $U(1)_Y$ factor of the 
standard model is, indeed, an effective manifestation of strongly interacting SU(2) gauge 
dynamics then the radiation history of the Universe needs some 
rewriting in the low redshift regime. This opens up the potential to 
explain the puzzling anomalies occurring for $z\le 20$. 
In addition, a near coincidence for the densities of cosmological dark 
matter ($\rho_{\tiny\mbox{DM}}\sim 0.3\rho_{\tiny\mbox{crit}}$) and dark energy 
($\rho_{\tiny\mbox{DE}}\sim 0.7\rho_{\tiny\mbox{crit}}$) is observed - a fact which, as we will 
argue below and have argued in \cite{Hofmann2005}, 
may be tightly related to strongly interacting, nonabelian gauge dynamics. 
(A slowly rolling Planck-scale axion receiving its mass through the topologically 
nontrivial fluctuations of SU(2)$_{\tiny\mbox{CMB}}$ - calorons, see \cite{frieman1995}.) 
The obvious because minimal candidate for such a scenario is a dynamical breaking of an 
SU(2) Yang-Mills theory down to its Abelian subgroup U(1). 

\section{SU(2) Yang-Mills thermodynamics}

\begin{figure}
\begin{center}
\leavevmode
\leavevmode
\vspace{5.3cm}
\includegraphics{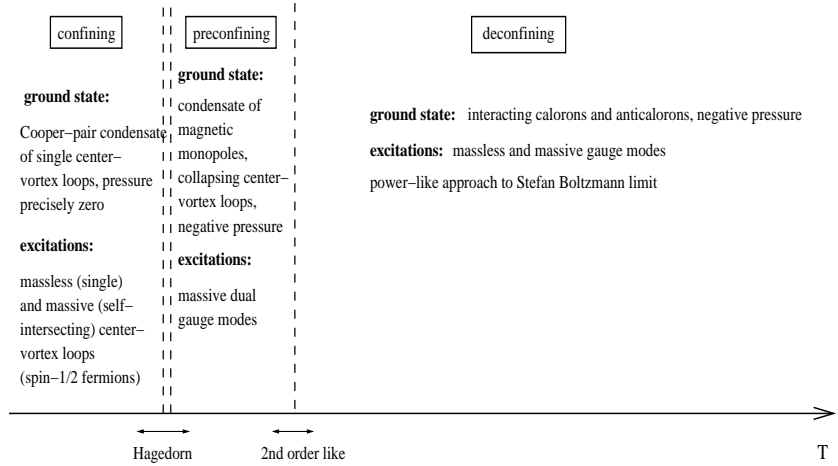}
\end{center}
\caption{\protect{The phase diagram of an SU(2) Yang-Mills theory.\label{Fig0}}}      
\end{figure}

A nonperturbative approach to SU(2) and SU(3) Yang-Mills 
thermodynamics was worked out recently 
in \cite{Hofmann2005,HerbstHofmann2004,HerbstHofmannRohrer2004}. 
This approach benefits from strong research efforts in SU(N) Yang-Mills theory 
performed both within the 
perturbative \cite{NP2004,Linde1980} and within the nonperturbative 
\cite{Polyakov1975,BPST1976,'tHooft1976,HarringtonShepard1977,GPY1981,Nahm1981,LeeLu1998,
KraanVanBaal1998,Ilgenfitz2002,Diakonov2004} realm. 
In \cite{Hofmann2005} we have derived the phase structure of an 
SU(2) and an SU(3) Yang-Mills theory. Each theory comes in three phases: 
a deconfining, a preconfing, and a confining one, 
see Fig.\,\ref{Fig0}. 
While the transition between the deconfining and the 
preconfining phase is of a second-order like nature the transition towards 
the confining phase is genuinely nonthermal and 
of the Hagedorn type. The latter transition goes with a change of statistics: 
The excitation in the preconfining phase are massive spin-1 bosons while they are massless and massive 
spin-1/2 fermions (single and selfintersecting center-vortex loops, respectively) 
in the confining phase. For our discussion of 
the CMB power spectra a large angles it is sufficient to resort to 
deconfining dynamics of the SU(2) theory. Therefore 
we will elucidate (but not derive) the dynamical facts 
for this phase only.  

\begin{figure}
\begin{center}
\leavevmode
\leavevmode
\vspace{5.7cm}
\includegraphics{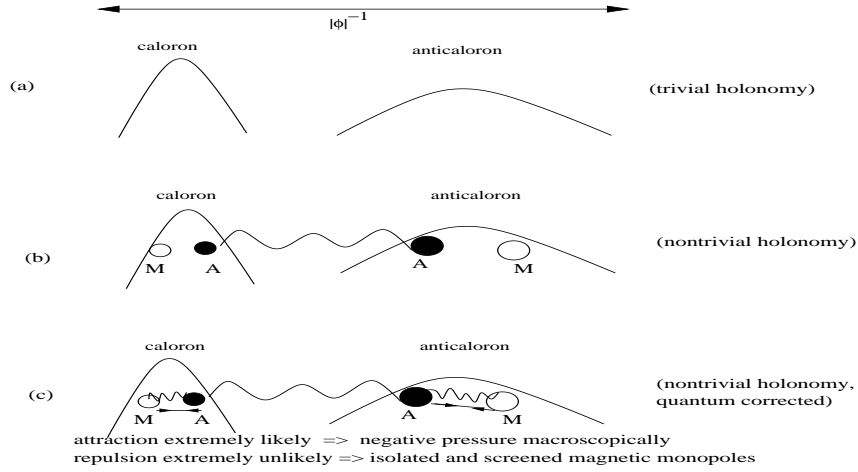}
\end{center}
\caption{\protect{Stepwise and selfconsistent derivation of a thermal ground state upon 
spatial coarse-graining. (a) A homogeneous and inert adjoint scalar field $\phi$ 
emerges from a pair of trivial-holonomy caloron and anticaloron upon spatial coarse-graining 
down to a resolution given by $|\phi|$. (b) Interactions between caloron and anticaloron, mediated by plane-waves of a 
resolution not much larger than $|\phi|$, induce a nontrivial holonomy and thus a BPS 
monopole-antimonopole pair in each configuration. (c) plane-wave fluctuations 
of resolution considerably larger than $|\phi|$ induce a potential between 
monopole and antimonopole which is attractive (repulsive) for a small (large) holonomy. \label{Fig1}}}      
\end{figure}

We start by considering a Euclidean formulation of the thermalized 
SU(2) theory where time is constrained to a 
circle, $0\le\tau\le\beta\equiv\frac{1}{T}$ (periodicity of gauge-field configurations). 
From a unique, nonlocal definition involving a pair 
of a noninteracting trivial-holonomy caloron and its anticaloron 
(Harrington-Shepard solution) \cite{HarringtonShepard1977} 
the computation of the $\tau$ dependence of the phase of a 
spatially homogeneous, quantum mechanically and statistically inert, 
adjoint scalar field $\phi$ is the first step in 
deriving a spatially coarse-grained action. 
Notice that it is consistent to determine the phase in terms 
of classical (Euclidean) field configurations since the periodic 
dependence in $\tau$ is solely determined by $T$, that is, 
dimensional transmutation is irrelevant for $\phi'$s phase. 
(To explain the term holonomy: A nontrivial holonomy is the feature of a finite-temperature gauge-field 
configuration that its Polyakov loop, evaluated at spatial 
infinity, is different from an element of the center $Z_2$ of SU(2). 
The Harrington-Shepard solution \cite{HarringtonShepard1977}, 
which is (anti)selfdual and thus energy-pressure free, is a periodic 
instanton in singular gauge with trivial holonomy (no substructure) and topological charge $\pm 1$. 
The Lee-Lu-Kraan-van-Baal solution \cite{LeeLu1998,KraanVanBaal1998} is 
(anti)selfdual with nontrivial holonomy and 
topological charge $\pm 1$. Because the nonvanishing $a_4(\vec{x}\to\infty,\tau)$ sets a 
mass scale proportional to temperature the solution exhibits a 
substructure in terms of a BPS magnetic monopole-antimonopole pair whose combined 
mass $M_m+M_a$ is $8\pi^2 T$. By computing the one-loop quantum weight for a 
nontrivial-holonomy caloron, which is a heroic deed, it can be shown 
that there is an attractive (repulsive), quantum-induced potential between monopole 
and antimonopole if the holonomy is small (large) \cite{Diakonov2004}.) 
Assuming the existence of a Yang-Mills scale $\Lambda_E$, which is strongly 
supported by one-loop perturbation theory \cite{NP2004}, the modulus $|\phi|$ 
follows. (At this level the scale $\Lambda_E$ only enters into a constraint for 
the finite size of the spatial volume that saturates the 
infinite-volume average.) Since $\phi$, describing averaged-over noninteracting trivial holonomy calorons and 
anticalorons at a spatial resolution $|\phi|$, turns out to be 
nondeformable it represents a fixed source to the coarse-grained 
Yang-Mills equations for the trivial-topology sector of 
the theory. Thus it turns out to be selfconsistent to consider (anti)caloron 
interactions, which are mediated by the topologically trivial sector, {\sl after} 
the spatial coarse-graining has been performed. These interactions manifest 
thermselves in terms of a pure-gauge configuration $a_\mu^{bg}$ solving the 
coarse-grained Yang-Mills equations. As a consequence, the pressure and the energy density 
of the ground state is shifted by (anti)caloron interactions 
from zero to $\mp 4\pi T\Lambda_E^3$. (The concept of a thermal 
ground state thus emerges in view of the average effect of interacting quantum fluctuations of 
trivial and nontrivial topology.) How this shift comes about 
on the microscopic level is depicted in Fig.\,\ref{Fig1}. 
On length scales not much 
smaller than $|\phi|^{-1}$ a gluon exchange between a trivial-holonomy 
caloron and its anticaloron essentially shifts the holonomy only, 
thereby creating a monopole-antimonopole pair in both the 
caloron and the anticaloron. Fluctuations of much higher resolution induce a potential 
between the monopole and its antimonopole. Since the latter attract for a 
small holonomy monopole and antimonopole eventually annihilate. 
Thus the quantum weight for the process of 
monopole-antimonopole creation and their subsequent annihilation essentially is 
given by that for a trivial-holonomy caloron \cite{GPY1981}. Depending 
on the scale parameter of the caloron this quantum weight can be sizable. 
In the opposite case of monopole-antimonopole repulsion (large holonomy) the (anti)caloron dissociates into 
an isolated but screened monopole and antimonopole. (The screening of magnetic charge 
is facilitated by intermediate, small-holonomy (anti)caloron fluctuations 
which generate short-lived magnetic dipoles.) The weight for such a process essentially is 
given by $\exp\left[-\frac{M_m+M_a}{T}\right]=\exp[-8\pi^2]$ 
which is an extremely small number. We thus conclude that the dissociation 
of (anti)calorons is an extremely rare process as compared to the fall-back process of 
a small-holonomy caloron onto trivial holonomy by monopole-antimonopole 
annihilation. The latter process involves {\sl attraction} between the (anti)caloron 
constituents: a situation which is responsible for the negative and 
spatially homogeneous ground-state pressure $P^{gs}=-4\pi T\Lambda_E^3$ emerging upon spatial coarse-graining. 
   
\begin{figure}
\begin{center}
\leavevmode
\leavevmode
\vspace{5.8cm}
\includegraphics{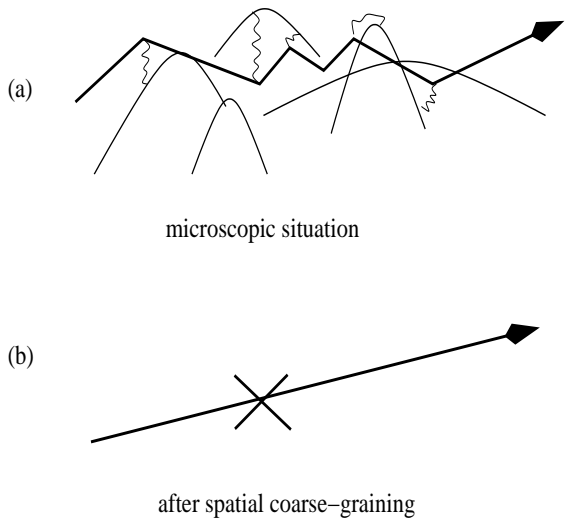}
\end{center}
\caption{\protect{Consecutive scattering of off-Cartan modes 
off calorons and anticalorons (a) and the emergence of mass 
by optimized spatial coarse-graining (b). \label{Fig2}}}      
\end{figure}

What about propagating excitations? Depending on the direction in color 
space an excitation either gets scattered off caloron or anticalorons (off-Cartan directions 
in a gauge where the spatial projection of an (anti)caloron is `combed' 
into a given color-space direction) or it propagates in an 
unadulterated way through the caloron-anticaloron 'forrest' (Cartan direction). 
In the former case an excitations describes a zig-zag like 
trajectory while there is straight-line propagation in the latter case. Upon spatial 
coarse-graining a zig-zag like propagation is converted into 
straight-line propagation but now subject to a mass term, 
see Fig.\,\ref{Fig2}. Straight-line propagation on the 
microscopic level is left untouched by spatial coarse-graining. 
After spatial coarse-graining the situation is summarized by 
the adjoint Higgs mechanism: two out of three 
directions in adjoint color space become massive 
($m_{W^\pm}=2e|\phi|=2e\sqrt{\frac{\Lambda_E}{2\pi T}}$ where $e$ denotes 
the {\sl effective} gauge coupling, the subscript $W^\pm$ solely indicates the massiveness and the electric 
charge of the excitation and is not to be associated with the massive bosons 
in the electroweak unification) due to coarse-grained, interacting (anti)calorons while the 
third direction remains massless ($m_\gamma=0$). Interactions between 
coarse-grained excitations are mediated by plane-wave fluctuations which, however, can not be further off 
their mass shell than $|\phi|^2$ since all fluctuations of resolution 
larger than $|\phi|$ are integrated into the pure-gauge configuration 
$a_\mu^{bg}$ already. This renders the effect of explicit interactions very 
small after spatial coarse-graining \cite{Hofmann2005,HerbstHofmannRohrer2004}. 
As far as large-angle signatures in the fluctuation map for the cosmic microwave background are 
concerned they do, however, play an important role. 

\begin{figure}
\begin{center}
\leavevmode
\leavevmode
\vspace{4.3cm}
\includegraphics{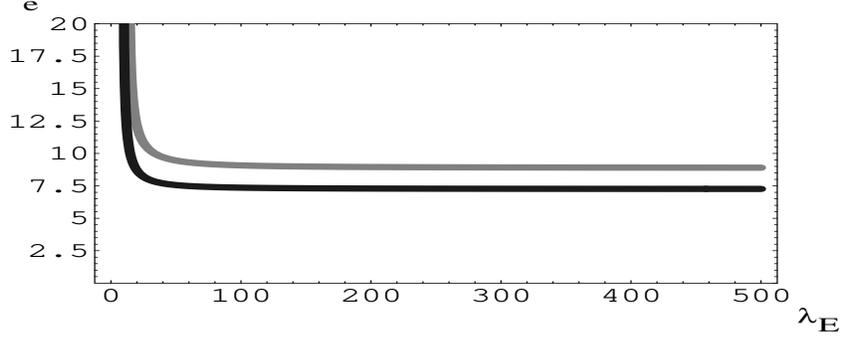}
\end{center}
\caption{\protect{The evolution of the effective gauge coupling $e$ with temperature in the deconfining phase. The grey
line depicts the SU(2) case while the black line is for SU(3). We have introduced a dimensionless temperature 
as $\lambda_E\equiv\frac{2\pi T}{\Lambda_E}$.
\label{Fig3}}}      
\end{figure}

The invariance of the Legendre transformations between thermodynamical 
quantities (thermodynamical selfconsistency \cite{Gorenstein1995}) demands a 
first-order evolution equation for the {\sl effective} gauge coupling $e$ \cite{Hofmann2005}. 
The solution of this equation is depicted in Fig.\,\ref{Fig3} for both SU(2) (grey line) and SU(3) 
(black line). A number of comments are in order (only discussing the SU(2) case). 
First, there is an attractor to 
the evolution being a plateau $e=8.89$ and a logarithmic divergence 
$e\propto -\log(\lambda_E-\lambda_{c,E})$ where $\lambda_{c,E}=13.867$. That is, the low-temperature 
behavior in the evolution of $e$ is independent of the boundary condition set at 
high temperatures. This is the celebrated ultraviolet-infrared decoupling property of the SU(2) 
Yang-Mills theory, which, in a logarithmic fashion, is already observed 
in perturbation theory. Second, the plateau value signals the conservation 
of the magnetic charge $g=\frac{4\pi}{e}$ of an isolated and screened 
magnetic monopole: Even after screening by intermediate small-holonomy caloron 
fluctuations each isolated monopole or antimonopole is a nonrelativistic 
particle for temperatures not too close to $\lambda_{c,E}$. The very limited mobility of a (anti)monopole then 
implies that the magnetic charge per unit volume is a conserved quantity. 
Third, only close to $\lambda_{c,E}$ do (anti)monopoles become mobile 
due to increased screening leading to the instability of (anti)calorons 
with respect to a switch to large holonomy (monopole condensation) \cite{Diakonov2004}: 
local charge conservation does no longer hold. This is the relevant regime for an investigation 
of polarization-temperature cross correlations in the angular power spectrum of the CMB. At 
$\lambda_{c,E}$ monopoles become precisely massless and thus Bose condense while the off-Cartan 
excitations exhibit a diverging mass and thus fall out of thermal equilibrium.   

\begin{figure}
\begin{center}
\leavevmode
\leavevmode
\vspace{4.3cm}
\includegraphics{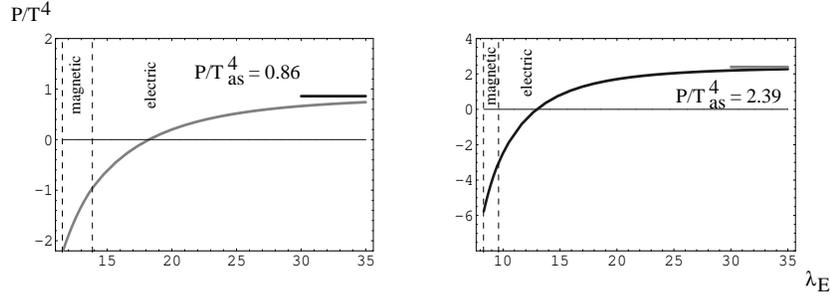}
\end{center}
\caption{\protect{The ratio $\frac{P}{T^4}$ for an SU(2) (left panel) and an SU(3) (right panel) Yang-Mills 
theory throughout its deconfining and preconfining
phase. \label{Fig4}}}      
\end{figure}

\begin{figure}
\begin{center}
\leavevmode
\leavevmode
\vspace{4.3cm}
\includegraphics{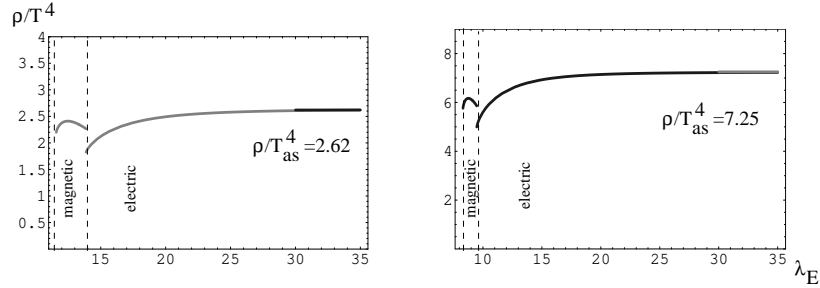}
\end{center}
\caption{\protect{The ratio $\frac{\rho}{T^4}$ for an SU(2) (left panel) and an SU(3) (right panel) Yang-Mills theory 
throughout its deconfining and preconfining
phase. \label{Fig5}}}      
\end{figure}

In Fig.\,\ref{Fig4} a plot of the total pressure $P$ over $T^4$ as a function of temperature 
is shown for both the deconfining and the preconfining phases. Notice the rapid (power-like) approach to 
the Stefan-Boltzmann limit. Notice also that the total pressure is negative shortly above $\lambda_{c,E}$ and even more
so in the preconfining phase. This is a consequence of the ever increasing dominance 
of the ground state when cooling the system in the regime where the temperature is comparable 
to the Yang-Mills scale and where excitations becoming increasingly massive. 
(Recall that the ground-state physics in the
deconfining phase originates from a spatial average over pairs of {\sl attracting}, {\sl annihilating}, 
and subsequently {\sl recreated} monopoles and antimonopoles while there are collapsing and recreated 
center-vortex loops in the preconfining phase making up the {\sl negative} ground-state pressure.)
In Fig.\,\ref{Fig5} we show a plot of the total energy density $\rho$ over $T^4$ as a 
function of temperature. Again, there is a power-like approach to 
the Stefan-Boltzmann limit. Notice the infinite-curvature 
dip at $\lambda_{c,E}$. The upwards jump toward the preconfining phase is a consequence 
of the dual gauge mode acquiring an extra polarization by 
the Abelian Higgs mechanism compared to the Cartan-excitation 
in the deconfining phase. To excite this additional polarization costs 
energy, therefore the jump. The steep slope to the right of the 
dip arises due to the logarithmic decoupling of 
the off-Cartan excitations when approaching 
$\lambda_{c,E}$. Because off-Cartan excitations possess infinite mass 
at $\lambda_{c,E}$ the Cartan excitation (our photon if the theory SU(2)$_{\tiny\mbox{CMB}}$ is 
considered) propagates in a completely unscreened way. Moreover, the photon is precisely 
massless because the magnetic coupling $g=\frac{4\pi}{e}$ 
to the monopole condensate vanishes at $\lambda_{c,E}$. This situation is strongly 
stabilized in terms of the dip in the energy density: Only a departure from thermal 
equilibrium, which is induced by an external source, 
will elevate the system into its preconfining phase. 

Let us briefly discuss how this likely happens in the real Universe. 
A Planck-scale axion field \cite{frieman1995}, which is spatially homogeneous 
on cosmological length scales and 
glued to the slope of its potential by cosmological friction at 
temperatures well above $\lambda_{c,E}$ (see e.g. \cite{Wilczek2004}), 
starts to roll for $\lambda\stackrel{>}\sim\lambda_{c,E}$. At a critical velocity 
of axion rolling, which is going to be reached eventually because the ratio of axion mass to the 
Hubble parameter increases with increasing axion velocity and because 
the bulk of the Universe's energy density is stored in the axion field, 
thermal equilibrium is sufficiently violated to overcome the 
discontinuity in the energy density of 
SU(2)$_{\tiny\mbox{CMB}}$ at $\lambda_{c,E}$. As a consequence, the 
photon will acquire a Meissner mass (visible superconductivity of 
the Universe's ground state).       

Taking the dual gauge 
boson mass as an `order parameter' for the second-order like transition at 
$\lambda_{c,E}$ (associated with an apparent gauge symmetry breaking $U(1)_D\to 1$ 
in the preconfining phase) we have determined the critical 
exponent to $\nu=0.5$ in \cite{Hofmann2005}. 
Even though both results, the pressure and the energy density, were obtained by a one-loop calculation they 
are accurate to within the 0.1\% level, see below.

\section{Electric-magnetic coincidence}

At the point $\lambda_{c,E}$ we encounter an exact coincidence in 
the electric and the magnetic description of the thermalized 
gauge system. Namely, no dynamical electric charges exist because the off-Cartan 
modes are decoupled and at the same time magnetic charges are 
condensed in such a way that they do not influence the properties 
of the left-over excitations (the magnetic coupling $g$ vanishes 
precisely at $\lambda_{c,E}$). The dynamical equations of the effective 
theory, which coincide with Maxwell's equations in the absence of sources, 
thus are invariant under an electric-magnetic duality transformation 
at the point $\lambda_{c,E}$. The ground state, although gravitationally 
detectable by a measurement of the expansion rate of the Universe, is not 
visible otherwise. To abandon the idea of a world ether, as it was done by Einstein a 
hundred years ago \cite{Einstein1905}, is correct as far as the electrodynamics of 
moving bodies is concerned but appears 
too radical when extending one's perspective by including gravity. 
This coins the name {\sl invisible ether} for today's SU(2)$_{\tiny\mbox{CMB}}$ 
monopole condensate not influencing the propagation of photon 
excitations (including those being excited by accelerated electric charges): 
At $\lambda_{c,E}$ the photon is massless and unscreened as we observe it today. 
Thus identifying the point 
$\lambda_{c,E}$ with the present temperature of the 
cosmic microwave background 
$T_{\tiny\mbox{CMB}}=2.351\times 10^{-4}\,$eV defines a 
boundary condition to the thermodynamics of the 
associated SU(2) Yang-Mills theory. As a result, we determine the 
scale of the Yang-Mills theory to $\Lambda_E=1.065\times 10^{-4}$\,eV. 
Knowing $\Lambda_E$ yields a prediction for the energy density $\rho^{gs}$ of the 
invisible ether (a contribution to dark energy). We have $\rho^{gs}=(2.444\times 10^{-4}\,\mbox{eV})^4$. 
If we take the measured value of today's density of dark 
energy to be $\sim (10^{-3}\,\mbox{eV})^4$ 
\cite{Perlmutter1998,Schmidt1998,WMAP2003I} then we derive that 
only about $0.36$\% of the Universe's present dark energy density is 
provided by the ground state of SU(2)$_{\tiny\mbox{CMB}}$. 
Therefore the bulk of today's dark energy density arises from another source. 
In \cite{Hofmann2005} we have scetched how a slowly-rolling Planck-scale 
axion besides generating today's dark energy density also may explain 
why there is a near coincidence of this value with that of the cosmological dark 
matter density. In addition, a Planck-scale axion, which becomes mobile 
whenever an SU(2) or an SU(3) Yang-Mills theory approaches its center (or confining) 
phase during the Universe's evolution and is frozen-in otherwise, represents a 
candidate mechanism for the generation of the observed baryon- 
and lepton asymmetries \cite{Hofmann2005}.  

The conceptually interesting thing is that Lorentz invariance, which is one of the defining 
features of the underlying Yang-Mills theory in the 
limit $T\to\infty$ and which is dynamically violated 
by interactions with a nontrivial, thermal ground state inside 
the deconfining and the preconfining phase 
(Higgs mechanism induced, temperature dependent masses), 
is {\sl dynamically} restored at the point $\lambda_{c,E}$ and in 
the effective theory at zero temperature (confining phase): A strongly 
interacting gauge theory restores an asymptotically valid 
spacetime symmetry at two specific points of its phase diagram.   

\section{Large-angle fluctuations in the CMB as radiative corrections}

We now would like to address how temperature fluctuations and temperature-polarization 
cross correlations at large angles may be generated in the associated power spectra 
of the cosmic microwave background. While we have something quantitative to say concerning 
the former case we need, for the time being, constrain ourselves to a 
qualitative statement in the latter case. 
\begin{figure}
\begin{center}
\leavevmode
\leavevmode
\vspace{2.7cm}
\includegraphics{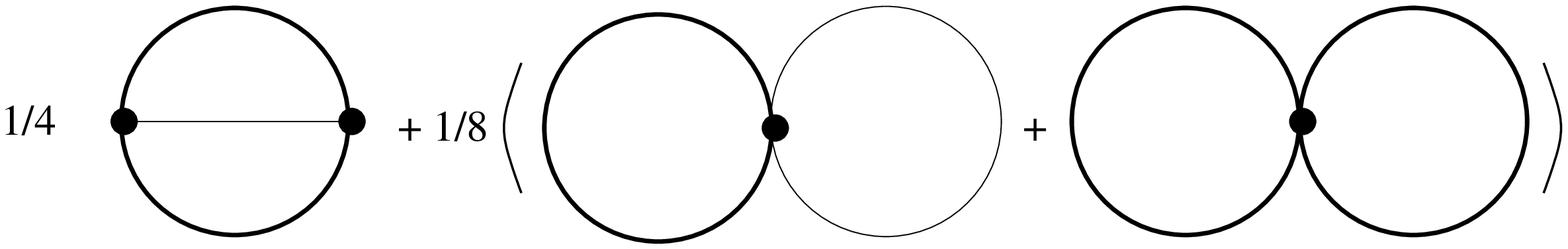}
\end{center}
\caption{\protect{Two-loop corrections to the pressure. 
The nonlocal diagram is the by-far dominating contribution. \label{Fig6}}}      
\end{figure}
Radiative corrections to the pressure at the two-loop level correspond to the set of diagrams as 
depicted in Fig.\,\ref{Fig6}. To evaluate these diagrams one has to work in a physical 
gauge (Coulomb-unitary) for a meaningful implementation of the cutoffs for the off-shellnes of 
quantum fluctuations. Recall that these cutoffs arise from the 
spatial coarse-graining inherent in the effective theory. The by-far dominating diagram 
is the nonlocal one. In Fig.\,\ref{Fig7} the relative 
correction to the free-gas pressure asrising from this diagram 
is shown as a function of temperature.    
\begin{figure}
\begin{center}
\leavevmode
\leavevmode
\vspace{4.2cm}
\includegraphics{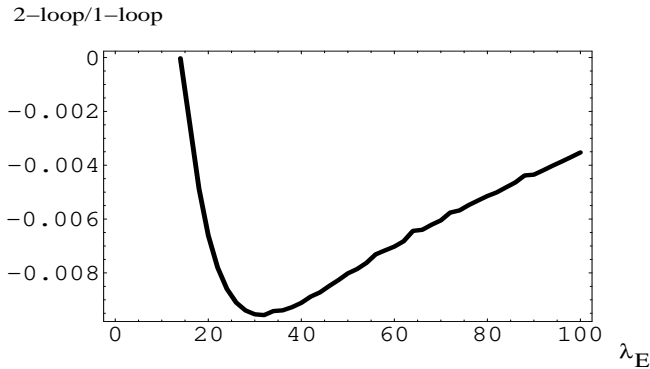}
\end{center}
\caption{\protect{The dominating two-loop correction to the 
pressure in the deconfining phase of an SU(2) Yang-Mills theory as a 
function of temperature.\label{Fig7}}}      
\end{figure}
The power of the dipole fluctuation measured 
in the CMB is $\left.\frac{\Delta T}{T_{\tiny\mbox{CMB}}}\right|_{l=1}=1.24\times 10^{-3}$ 
which is close to the value obtained from the modification of 
black-body spectra, see \cite{SHG2006}.  

\begin{figure}
\begin{center}
\leavevmode
\leavevmode
\vspace{4.3cm}
\includegraphics{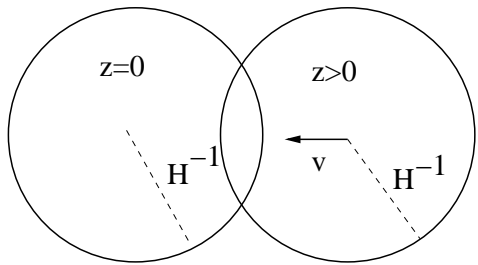}
\end{center}
\caption{\protect{Motion of a Hubble volume from $z\sim 3$ to $z=0$ 
in a de Sitter Universe.\label{Fig8}}}      
\end{figure}

The standard explanation for the dipole moment is in terms of a 
net velocity $v\sim 370$\,km s$^{-1}$ of the solar rest frame as compared 
to the CMB rest frame \cite{PeeblesWilkinson1968}. This is a purely 
kinematical explanation of the dipole in terms of the relativistic 
Doppler effect. It is known for a long time that the inferred velocity 
$v$ is unexpectedly larger than the relative velocity measured between the center-of-mass 
frame of galaxies and the solar system, see for example 
http://apod.gsfc.nasa.gov/apod/ap990627.html. It is well possible 
that a combination of kinematical and dynamical effects 
generates the dipole as follows: Let us, for simplicity, imagine a pure 
de Sitter Universe such that the Hubble radius, which sets the size of the horizon, 
is constant in time. Assume that the solar system started to move at $z\sim 3$ where 
global temperature fluctuations peak. Due to its net velocity with respect to the comoving frame 
the solar system's horizon volume at $z=0$ is spatially shifted as compared 
to that at $z\sim 3$. While a temperature fluctuation is of horizon-size 
at $z\sim 3$ temperature fluctuations no longer are global at $z=0$ because the latter 
horizon volume has picked up a sequence of temperature changes 
along the direction of its motion by diving into 
formerly causally disconnected regions, 
see Fig.\,\ref{Fig8}. As a consequence, the pure kinematically inferred 
$v\sim 370$\,km s$^{-1}$ would represent an upper bound only, 
the actual value may be signifantly lower. 
In reality there is no pure de Sitter expansion at $0\le z\stackrel{\sim}< 3$ 
but this does not alter the qualitative validity of the argument. 

Let us now discuss how a large cross correlation between temperature 
fluctuation and electric field polarization at large angles is likely to be generated 
without relying on the hypothesis of an early 
reionization of the Universe. Fig.\,\ref{Fig7} shows that the dominating 
radiative correction to the photon-gas pressure starts to become sizable at 
$\lambda_{E}\sim 10\,\lambda_{c,E}$. In this regime the isolated 
magnetic monopoles, which are electrically charged with respect to U(1)$_{Y}$, 
become increasingly light by an ever increasing screening by intermediate 
small-holonomy caloron fluctuations. This renders them explicit 
and mobile electric charges capable of amplifying a primordially 
existing cross correlation. At $\lambda_{c,E}$ monopoles condense 
and therefore are not available as isolated scattering centers 
anymore.  

Finally, we would like to stress that it may be premature to take the observed 
large-angle anomalies, as suggested by the analysis of the one-year WMAP data, 
at face value as far as their cosmological origin is concerned \cite{Copi2005}.      

\section{Summary, conclusions and future work}

We have proposed that a strongly interacting SU(2) pure gauge theory (SU(2)$_{\tiny\mbox{CMB}}$) 
of Yang-Mills scale $\Lambda_E=1.177\times 10^{-4}$\,eV masquerades 
as the U(1)$_Y$ factor of the standard model of particle physics 
within the present cosmological epoch. This proposal looks in so far 
viable and consistent as (i) there exists a dynamical stabilization 
mechanism for the exact restoration of Lorentz invariance at a 
particular point in the phase diagram of the Yang-Mills theory (the boundary between the 
deconfining and the preconfining phase), (ii) the ground-state energy density of 
SU(2)$_{\tiny\mbox{CMB}}$ (not coupling the Planck-scale axion to it) at this point represents 
only about $0.36$\% of the measured density in 
dark energy of the present Universe (assuming $\rho_\Lambda\sim (10^{-3}\,eV)^4$) , (iii) the dipole strength in the temperature 
map of the CMB is close to the effect arising from nonabelian fluctuations in SU(2)$_{\tiny\mbox{CMB}}$, 
(iv) there is a mechanism for providing a large correlation between 
temperature fluctuation and electric field polarization at large angles in terms of mobile and isolated 
electrically charged monopoles (the hypothesis of an early 
reionization may turn out to be redundant), and (v) coupling a (slowly rolling) 
Planck-scale axion to the theory, possibly explains the 
observed near coincidence between cosmological dark matter and 
dark energy. (Notice, however, that this would imply that structure 
formation would be due to ripples and lumps in the 
coherent axion field \cite{Wetterich2001}.) The increasing rate of rolling of the latter will 
eventually destroy the present thermal equilibirum and elevate 
SU(2)$_{\tiny\mbox{CMB}}$ into its preconfining phase where the 
photon is Meissner massive.  

Furthermore, the system SU(2)$_{\tiny\mbox{CMB}}$ plus 
Planck-scale axion may provide a future theoretical framework to 
investigate the overall strength and distribution of intergalactic 
magnetic fields. (For a slight deviation from thermal equilibrium patches of the 
Universe's ground state are visibly superconducting by the 
condensate of electric monopoles coupling to its excitations.)

To substantiate the scenario further we need to investigate various angular 
two-point correlations by a diagrammatic analysis of the radiative 
corrections in the deconfining phase of SU(2)$_{\tiny\mbox{CMB}}$. 
On a microscopic level, we also may 
investigate CMB photon scattering processes off individual, electrically 
charged monopoles for $T>T_{\tiny\mbox{CMB}}$. (At a given temperature 
$T>T_{\tiny\mbox{CMB}}$ the number density, the mass, and the charge radius 
of the latter can be reliably estimated \cite{Hofmann2005,Diakonov2004}.) 
This provides a handle on the amount of induced electric polarization. 
A fluctuating Planck-scale 
axion should introduce an asymmetry between the 
electric and the magnetic polarization-temperature 
cross correlation at large angles. It also should make the expectation in 
the large-angle fluctuation of electric times magnetic mode nonvanishing. 
If future observations of the CMB at large angles detect a clear 
signal for CP violation then this would be another indication 
that the system SU(2)$_{\tiny\mbox{CMB}}$ plus 
Planck-scale axion is responsible for the ground-state 
physics of our present Universe.

\end{document}